\documentclass[review]{elsarticle}

\usepackage{lineno,hyperref}
\usepackage{graphicx}   % Include figure files
\usepackage{bm}% bold math
\usepackage{amsfonts}
\usepackage{epsfig}
\usepackage{epstopdf}
\usepackage{amsmath}
\usepackage{float}
\usepackage{braket}
\usepackage[all]{xy}

\usepackage{lineno,hyperref}
\usepackage{graphicx}   % Include figure files
\usepackage{bm}% bold math
\usepackage{amsfonts}
\usepackage{epsfig}
\usepackage{epstopdf}
\usepackage{amsmath}
\usepackage{float}
\usepackage{braket}
\usepackage{listings}
\usepackage{color}
\usepackage[all]{xy}

\modulolinenumbers[5]

\journal{Journal of Molecular Graphics and Modelling}

\begin{document}

\begin{frontmatter}

\title{GPView: A Program for Wave Function Analysis and Visualization}

\author[mymainaddress]{Tian Shi\corref{mycorrespondingauthor}}
\cortext[mycorrespondingauthor]{Corresponding author}
\ead{gpview@life-tp.com}

\author[mysecondaryaddress]{Ping Wang}
\ead{wangpinggl@gmail.com}

\address[mymainaddress]{Department of Chemistry, Wayne State University, Detroit, Michigan 48202, United States}
\address[mysecondaryaddress]{Department of Computer Science, Wayne State University, Detroit, MI 48202, United States}

\begin{abstract}
In this manuscript, we will introduce a recently developed program GPView, which can be used for wave function analysis and visualization. The wave function analysis module can calculate and generate 3D cubes for various types of molecular orbitals and electron density of electronic excited states, such as natural orbitals, natural transition orbitals, natural difference orbitals, hole-particle density, detachment-attachment density and transition density. The visualization module of GPView can display molecular and electronic (iso-surfaces) structures. It is also able to animate single trajectories of molecular dynamics and non-adiabatic excited state molecular dynamics using the data stored in existing files. There are also other utilities to extract and process the output of quantum chemistry calculations. The GPView provides full graphic user interface (GUI), so it very easy to use.
It is available from website \href{http://life-tp.com/gpview}{http://life-tp.com/gpview}.
\end{abstract}


\end{frontmatter}

%\linenumbers

\section{Introduction}

Quantum chemistry calculations has produced lots of data, such as energy, dipole, force, wave functions, density matrices and so on.
Some of them are widely used to simulate experimental results and derive novel theoretical methods. 
Other data may need to be analyzed and visualized by some professional software, 
such as GaussView, 
MOLDEN \cite{molden}, 
multiwfn \cite{lu2012multiwfn},
TheoDORE\cite{plasser2014new1,plasser2014new2} and NancyEX \cite{etienne2014toward}. 
For instance, parameters of real space wave functions are stored in Gaussian 09 \cite{Gaussian} fchk files.
We have to perform further calculations and convert them to 3D cubes, in order to view
molecular orbitals and electron density by graphic software.
However, there are still many data going to the trash before we use them. 
A good example is on rwf files produced by Gaussian 09 \cite{Gaussian}. 
Since their sizes are large, they are deleted immediately after QC calculations unless users tell the program not to.
There are several reasons for throwing away these data. 
First, they may be irrelevant to experimental results.
Second, programs that can make use of these data and produce meaningful results have not been developed yet.
Third, some software require knowledge of python, perl or shell script, and may need to work with third party libraries, therefore they are only used by a small community of computational chemists. 
For instance, transition density matrices \cite{Tretiak2002-CR-1} (TDMs) play a crucial role in the Exciton Scattering approach \cite{Wu2008-PRL-1,Wu2006-NP-1}, since electronic excited states belonging to different exciton bands are collected based on the color maps or contour plots of TDMs \cite{Wu2006-NP-1}. 
However, it is difficult to extract TDMs in the basis of atomic orbitals from QC outputs and to calculate contracted TDMs \cite{Tretiak2002-CR-1} in the basis of atoms. 
Moreover, although different methods for analyzing electronic excitations have been summarized in \cite{plasser2014new1} and \cite{plasser2014new2}, there is no easy-to-use software that implements all of these state-of-the-art methods.

Therefore, we recently develop a software, named GPView, to do wave function analysis and visualization for electronic excitations. 
The analysis module, including wave function and density matrix based analysis, allows us to
study excited states by different methods \cite{plasser2014new1,Tretiak2002-CR-1}.
The results can be viewed by the visualization module.
The GPView program provides full graphic user interface (GUI),
which makes it easy to use, even for those who without programing experience.

The rest of this chapter is organized as follows. 
The underlining theoretical background of analysis module is introduced in section \ref{methodology}. 
It is followed by introducing visualization and analysis functions in section \ref{program}. 
Finally, in section \ref{application}, we will apply different methods to a model molecule.

\section{Theoretical methodology}
\label{methodology}
GPView is a visualization tool, which can be used to visualize molecular and electronic structures. 
It is also an analysis tool, which is capable to calculate and generate 3D grids for molecular
orbitals and electron density. In this section, the theoretical
background behind these functions are briefly introduced. 
We will start from molecular orbitals (MO) and electron density.
Then, in the following subsections, we will outline analysis methods related with electronic excitations.
The technical details and derivations will not be explained, 
since you can always find them in these books \cite{helgaker2014molecular,szabo1989modern} and papers 
\cite{plasser2014new1,plasser2014new2,plasser2012analysis,martin2003natural,head1995analysis,Tretiak2002-CR-1}.

\subsection{Molecular Orbitals (MOs)}
Molecular orbitals are linear combination of atomic orbitals (LCAO), i.e.
\begin{equation}
	\phi_i(r)=\sum_j C_{i,j}\chi_{j}(r),
\end{equation}
where $\phi_i(r)$ and $\chi_j(r)$ represent the wave functions of molecular and atomic orbitals, respectively. 
The expansion coefficients $C_{i,j}$ are called molecular orbital coefficients.

Given wave functions $\phi_i(r)$ and occupation numbers $\zeta_i$ of occupied MOs, the total SCF density is expressed as
\begin{equation}
	\rho(r)=\sum_i^{N_{occ}}\zeta_i(\phi_i(r))^2,
\end{equation}
where the occupation number $\zeta_i=2$ for restricted closed shell system. 

\subsection{Transition Density Matrices (TDMs)}
The approaches in this section are based on the analysis of one-electron transition density (TDM)
matrices \cite{Tretiak2002-CR-1,plasser2012analysis}.
\begin{equation}
	T^\alpha_{mn}=\braket{\Phi^\alpha_m|a_m^\dagger a_n|\Phi^0_n},
\end{equation}
where $\Phi^0_n$ and $\Phi^\alpha_m$ represent the ground state and excited state $\alpha$. 
The subscript $m$ and $n$ denote atomic orbitals, in other words atomic basis functions. 
$T_{mm}^\alpha$ represents the net charge induced on the $m$-th atomic orbital when 
there is an electronic transition between ground state and excited state $\alpha$, while $T_{mn}^\alpha$ ($m\neq n$) represents the joint amplitude of finding an extra electron on orbital $m$ and a
hole on orbital $n$ \cite{Tretiak2002-CR-1}.

To visualize electronic transition modes, matrix elements belong 
the same atom are contracted to a single element. In the GPView, the following two approach are adopted for the contractions
\begin{equation}
	\label{tdmc-1}
	\Omega_{AB}^\alpha=\sqrt{\sum_{m\in A, n\in B} |T_{mn}^\alpha|^2},
\end{equation}
and
\begin{equation}
	\label{tdmc-2}
	\Omega_{AB}^\alpha=\sum_{m\in A, n\in B} |T_{mn}^\alpha|,
\end{equation}
where $A$ and $B$ are labels for atoms. The third approach is based on the concept of charge 
transfer number suggested by Plasser and Lischka in \cite{plasser2012analysis},
\begin{equation}
	\label{ctnm-1}
	\Omega_{AB}^\alpha=\sum_{m\in A, n\in B} (T^\alpha S)_{mn}(S T^\alpha)_{mn},
\end{equation}
where $T^\alpha$ and $S$ denote the transition density matrix and overlap matrix.
This definition is analogous to Mayer's bond order \cite{plasser2012analysis}.
Here, we name the matrix obtained from equation \ref{tdmc-1} and \ref{tdmc-2} as 
{\it contracted transition density matrix (CTDM)} and from equation \ref{ctnm-1} as {\it charge transfer number matrix (CTNM)} in GPView. 

It is worth to mention that the CTNMs can be further used 
to calculate the charge transfer numbers between different molecular fragments 
by simply taking the sum of corresponding matrix elements. 
For visualization purpose, we can calculate the square roots of absolute values of matrix elements (SCTNM). The SCTNM can be displayed as color maps or contour plots. 
As to the population analysis, the CTNM can be used to 
calculate the partition ratio, coherence length and so on \cite{plasser2012analysis}. 
Compared with the original TDM, the CTDM and CTNM are much smaller in size, which make them easy to be stored and shared.

\subsection{Natural Transition Orbitals (NTOs)}
Natural Transition Orbitals have become a standard tool to study the 
optically electronic transitions \cite{martin2003natural,surjan2007natural}.
They give electron-hole pictures of electronic excitations.
They are obtained by diagonalizing TDMs in 
the basis of canonical molecular orbitals. 
The matrix elements $T^{\alpha, MO}_{mn}$ represent
the joint amplitude of electronic transitions between molecular orbitals, 
instead of atomic basis functions.
In the case of CIS and TDA \cite{hirata1999time} approximation, they are the same as
CI-coefficients promoted from MO $\phi_n$ to $\phi_m$ \cite{plasser2014new1}. 
It is well known that TDMs are not symmetric. 
They may not be squared matrices, since the number of occupied and virtual orbitals are different. Therefore, we can not diagonalize TDMs.

Suppose we have an $M\times N$ transition density matrix $T^{\alpha,MO}$, 
by doing Singular Value Decomposition (SVD),
\begin{equation}
	T^{\alpha,MO}=U\Lambda V^{tr},
\end{equation}
where $\Lambda$ is  a diagonal matrix with $\Lambda_{ii}=\sqrt{\lambda_i}$ ($i=1,\dots,M$) represent eigenvalues. 
The matrix $U$ and $V$ store eigenvectors (i.e. expansion coefficients) for hole and particle NTOs, respectively. It is worth to mention that $U$ and $V$ are eigenvectors of matrix $TT^{tr}$ ($M\times M$) 
and $T^{tr}T$ ($N\times N$).
Regardless frozen orbitals, $M$ and $N$ ($M<N$) are equal to the number occupied orbitals and 
virtual orbitals, respectively.

With matrix $U$ and $V$, we can define hole and particle NTOs \cite{martin2003natural,dreuw2005single,plasser2014new1} as
\begin{eqnarray}
	\psi^h_j(r)=\sum_i U_{ij}\phi^o_i(r), \\
	\psi^p_j(r)=\sum_i V_{ij}\phi^v_i(r),
\end{eqnarray}
where $\phi^o$ and $\phi^v$ represent occupied and virtual molecular orbitals, respectively. The wave functions for hole NTOs $\psi_j^h(r)$ and particle NTOs $\psi_j^p(r)$ are associated to the same eigenvalue $\sqrt{\lambda_j}$. Since $M<N$, the original $M\times N$ matrix is reduced to an $M\times M$ matrix in the basis of NTOs.

With NTOs, the hole density $\rho^h(r)$, particle density $\rho^p(r)$ and transition density $\rho^t(r)$ are expressed as
\begin{eqnarray}
	\rho^h(r)=\sum_j \lambda_j(\psi^h_j(r))^2, \\
	\rho^p(r)=\sum_j \lambda_j(\psi^p_j(r))^2, \\
	\label{TransitionDensity}
	\rho^t(r)=\sum_j \sqrt{\lambda_j}\psi^h_j(r)\psi^p_j(r),
\end{eqnarray}
where in equation \ref{TransitionDensity}, the hole and particle NTOs are rearranged, 
so that both $\psi^h_j(r)$ and $\psi^p_j(r)$ correspond to the same eigenvalue $\sqrt{\lambda_j}$. 
The hole, particle and transition density can also be defined by NTO with the largest eigenvalue (close to 1).

Finally, the natural transition orbital partition ratio \cite{plasser2012analysis} is defined as
\begin{equation}
	PR_{NTO}=\frac{(\sum_i \lambda_i)^2}{\sum_i \lambda_i^2}.
\end{equation}

\subsection{Natural Orbitals (NOs)}
The Natural Orbitals are obtained by diagonalizing the state density matrices. 
Particularly, we diagonalize CI density matrices to get NOs for different 
excited states in GPView. 
Since a state density matrix is symmetric, 
it can be diagonalized by an unitary matrix $U$,
\begin{equation}
	D^\alpha = U \Lambda U^{tr},
\end{equation}
where the $\Lambda$ is a diagonal matrix with $\Lambda_{ii}=n_i$ are eigenvalues.
$n_i$ is also named as occupation number. 
The number of electrons in a system is identical to the sum of occupation numbers
\cite{plasser2014new1},
\begin{equation}
	n=\sum_i n_i.
\end{equation}
With matrix $U$, NOs are defined as
\begin{equation}
	\label{lcmo-no}
	\psi_j(r)=\sum_i U_{ij}\phi_i(r),
\end{equation}
where $\phi_i(r)$ represent canonical molecular orbitals. 
Similarly, we can define the total density 
as the weighted sum over squared NOs,
\begin{equation}
	\rho^\alpha(r)=\sum_i n_i(\psi_i^\alpha(r))^2,
\end{equation}
where $\alpha$ denotes excited state $\alpha$. We can take the difference between
density of state $\alpha$ and ground state
\begin{equation}
	\Delta\rho(r)=\rho^\alpha(r)-\rho(r),
\end{equation}
to get the difference density. The capability and limitation of using difference density to study electronic transition can be found in \cite{dreuw2005single}.

\subsection{Natural Difference Orbitals (NDOs)}
The natural difference orbitals are used to describe the electron attachment and detachment processes.
They are based on difference of density matrices,
\begin{equation}
	\Delta^{0\alpha}=D^{\alpha}-D^{0},
\end{equation}
where $D^0$ and $D^\alpha$ are density  matrix for ground and $\alpha$ excited state. Since both of them are symmetric, $\Delta^{0\alpha}$ is also a symmetric matrix.
Therefore, it can be diagonalized
\begin{equation}
	\Delta^{0\alpha}=U\Lambda U^{tr},
\end{equation}
where eigenvalues $\Lambda_{ii}=\kappa_i$. 
Similar to equation \ref{lcmo-no}, we can get the NDOs, which can be classified into two groups according to signs of their eigenvalues, 
where negative and positive describe electron detachment and attachment processes, respectively. Mathematically, we need to define two vectors $d$ and $a$ of eigenvalues
\begin{eqnarray}
	d_i=-\min(\kappa_i,0),\\
	a_i=\max(\kappa_i,0).
\end{eqnarray}
The detachment and attachment densities are then defined as the weighted sum of squared NDOs \cite{head1995analysis,plasser2014new1,dreuw2005single}
\begin{eqnarray}
	\rho^d(r)=\sum_i d_i (\psi_i^d(r))^2, \\
	\rho^a(r)=\sum_i a_i (\psi_i^a(r))^2.
\end{eqnarray}

At CIS level, NTOs are composed of pairs of virtual and
occupied orbitals (or electron-hole pairs), while NOs
describe individual unpaired electrons, and NDOs refer to
independent attachment and detachment contributions \cite{plasser2014new1}.

It is worth to mention that overlap matrix is positive defined \cite{szabo1989modern}. If the eigenvalues are negative, the overlap matrix is unphysical. In this situation, $S^{\frac{1}{2}}$ can not be obtained, so that you can not use GPView to calculate NOs and NDOs.

\section{Main functions}
\label{program}
GPView is a software for visualization and analysis purpose.
The full graphic user interface makes it user friendly.  
We will briefly introduce the main functions in the following two subsections. 
More detailed descriptions can be found in the user's manual on
\href{http://life-tp.com/gpview}{http://life-tp.com/gpview}.

\subsection{Visualization module}
\begin{enumerate}
	\item Visualize molecular structures. 
	GPView can display molecular structures as balls and sticks and save them as figures. 
	It supports many file formats as input. 
	For example, it can read structures from Gaussian input, output, fchk and wfx files. 
	It can also read structures from PDB and XYZ files. The molecules can be displayed as balls and sticks, liquorice balls and sticks, VDW spheres, and protein ribbon bands, typically. All screen shots can be saved as figures into PNG files.

	\item Animate a single trajectory of molecular dynamics (MD).
	During the MD for small molecules, molecular structures are stored in the XYZ files. 
	GPView can read structures from XYZ files, create animations and save screen shots as videos (AVI) and figures (PNG).
	
	\item Animate a single trajectory of non-adiabatic excited state molecular dynamics (NA-ESMD). 
	This function is useful for visualize and analyze results of NA-ESMD \cite{barbatti2014newton,du2015fly}. 
	In addition to molecular structures, we also want to keep track of several other variables, such as potential energy surfaces, state populations, non-adiabatic coupling terms, bond length and hopping probabilities. 
	They are all stored in TXT files with special format. This function is tested by the output of GPV-ESMD which will be introduced elsewhere.

	\item Visualize electronic structures (Fig.\ref{fig:view-cube}). 
	GPView can interface with Gaussian Cube files, which store grid points of 3D objects, and show the iso-surfaces for molecular orbitals and electron density. 
	There are several options related with properties of iso-surfaces, so that we can easily choose
	values, contours and colors of the displayed surfaces.

	\item Visualize a pair of 3D objects. 
	Sometimes, we may need to view and compare two objects, such as HOMO-LUMO, hole-particle denisty and detachment-attachment density. 
	The 'visualize a pair of cubes' module in GPView makes it simple, because we are able to load two objects into the same window, and interact with them simultaneously.
	%----------------------------------------------------------------------
	\begin{figure}[H]
		\begin{center}
			\includegraphics[width=1.0\textwidth]{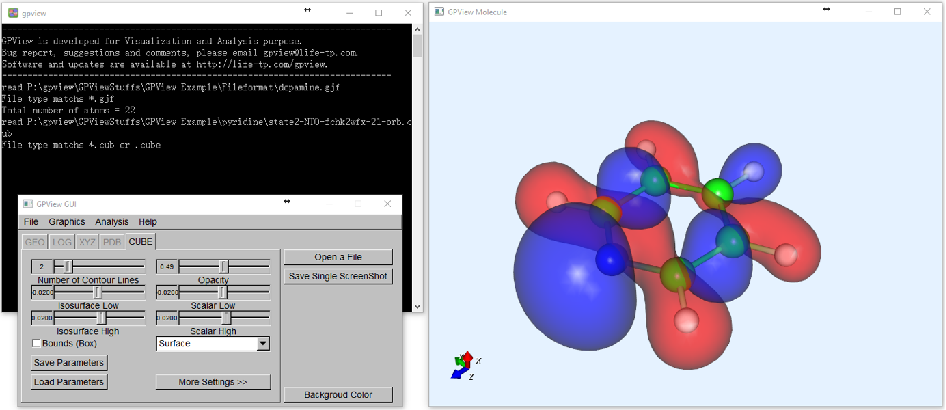}
		\end{center}
		\vspace{-0.5cm}
		\caption{Visualization of molecular orbitals and electron density.
			\label{fig:view-cube}
		}
	\end{figure}
	%----------------------------------------------------------------------
	
	\vspace{-0.1in}
	\item Contour plots and color maps of matrices (Fig.\ref{fig:view-matrix}).
	This function is related with visualization of excited state electronic modes based on CTDMs and SCTNMs.
	It has found applications in classifying excited states to different exciton bands \cite{Wu2008-PRL-1,Wu2006-NP-1}, and distinguishing excitonic and charge transfer states \cite{plasser2012analysis,Tretiak2002-CR-1,huang2015theoretical} for organic conjugated molecules.
	GPView can read these matrices from TXT files and show them as 2D contours or color maps.
	%----------------------------------------------------------------------
	\begin{figure}[H]
		\begin{center}
			\includegraphics[width=1.0\textwidth]{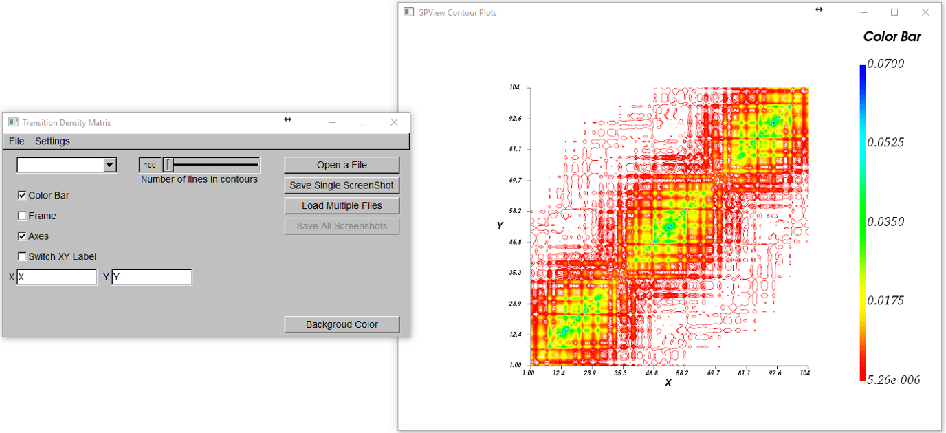}
		\end{center}
		\vspace{-0.5cm}
		\caption{Contours and color maps of TDMs.}
		\label{fig:view-matrix}			
	\end{figure}
	%----------------------------------------------------------------------
	\vspace{-0.1in}
	\item Plot density of states and spectrum. 
	We can use GPView to read energy and oscillator strength from Gaussian output or TXT files, and plot the density of states and spectrum. There are two choices for broaden functions: Gaussian and Lorentzian.
\end{enumerate}

\subsection{Analysis module}
\begin{enumerate}
	\item Interface with Gaussian rwf dumped files. 
	During quantum chemistry calculations, Gaussian 09 \cite{Gaussian} outputs many useful matrices to rwf files, which can be transfered into TXT files by the 'rwfdump' utility.
	Then, GPView can be applied to interface these files and extract them to TXT files in matrix (or half matrix) format.
	These matrices include overlap matrix, MO coefficients, density matrices, TDMs, CI-coefficients and so on.
%	\vspace{-0.1in}
	\item Calculate CTDM and CTNM. With the TDM and overlap matrix, we can calculate the CTDM and CTNM by GPView. These matrices can be visualized as contours or color maps, representing electronic modes of excited states. As discussed in theoretical methodology section, the CTNM can be further used in population analysis.
%	\vspace{-0.1in}
	\item Real space wave functions. GPView can read parameters from Gaussian 09 output, fchk and rwf dumped files, and then perform calculations and generate 3D cubes for molecular orbitals and electron density. This procedure has two steps. First, we can use GPView to interface with fchk files and calculate molecular orbitals, natural transition orbitals, natural orbitals or natural difference orbitals. The parameters will be stored in wfx files. Second, we will use GPView to interface with wfx files, and then calculate orbitals and electron density. The 3D grid points will be generated and written into cube files.
\end{enumerate}

\section{GPView Implementation}
\label{sec:program}
The GPView project started from early 2015 and the first version of the package was released in January 2016.
GPView is written in C++ programming language, and it depends on three other free, cross-platform and open-source libraries.
The Armadillo C++ linear algebra library\footnote{http://arma.sourceforge.net} 
is used for matrix and vector calculations.
The state-of-the-art Visualization ToolKit (VTK) library\footnote{http://www.vtk.org} provides a powerful tool to
conduct real-time three-dimensional rendering for molecular structures (balls and sticks)
and electronic structures (orbitals and density are represented by iso-surfaces).
The Graphic User Interface is realized by calling the Fast Light ToolKit (FLTK) library\footnote{http://www.fltk.org}.

The package of GPView has been tested on Windows 7/8/10 and Mac OS X systems. The installers are available on our website\footnote{http://life-tp.com/gpview}.
It should be noted that we don't need to install these libraries before running the GPView program since all the dependent libraries have been preloaded in the installers, which is very convenient for the users.

\section{Applications}
\label{application}
\begin{figure}[H]
	\begin{center}
		\includegraphics[width=0.9\textwidth]{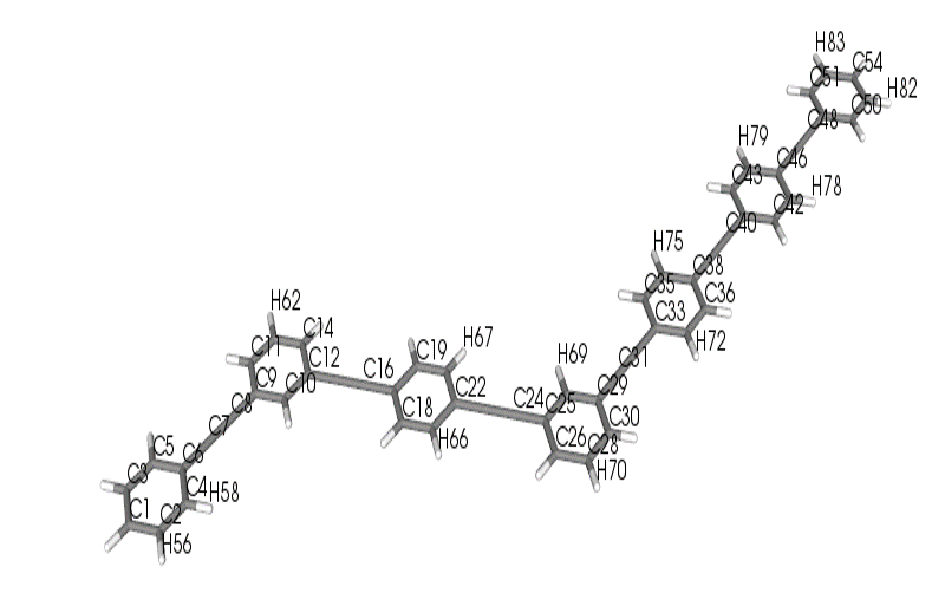}
	\end{center}
	\vspace{-0.5cm}
	\caption{The model conjugated molecule.
		\label{molecule}
	}
\end{figure}

In this section, we will illustrate the capability of GPView in wave function analysis of electronic excitations.
Quantum chemistry calculations have been performed on a branched dendritic phenylacetylene (PA) based molecule, which is composed of two-, three- and four- ring linear poly(phenylene ethynylene) (PPE) units linked through meta-substitutions \cite{soler2012analysis} (see Fig.~\ref{molecule}). The molecule is optimized at CAM-B3LYP/6-31G level of theory \cite{Yanai2004-CPL-1}, followed by TDDFT calculations at the same level of theory for 10 singlet excited states by Gaussian 09 \cite{Gaussian}.

\begin{figure}[H]
	\begin{center}
		\includegraphics[width=1.0\textwidth]{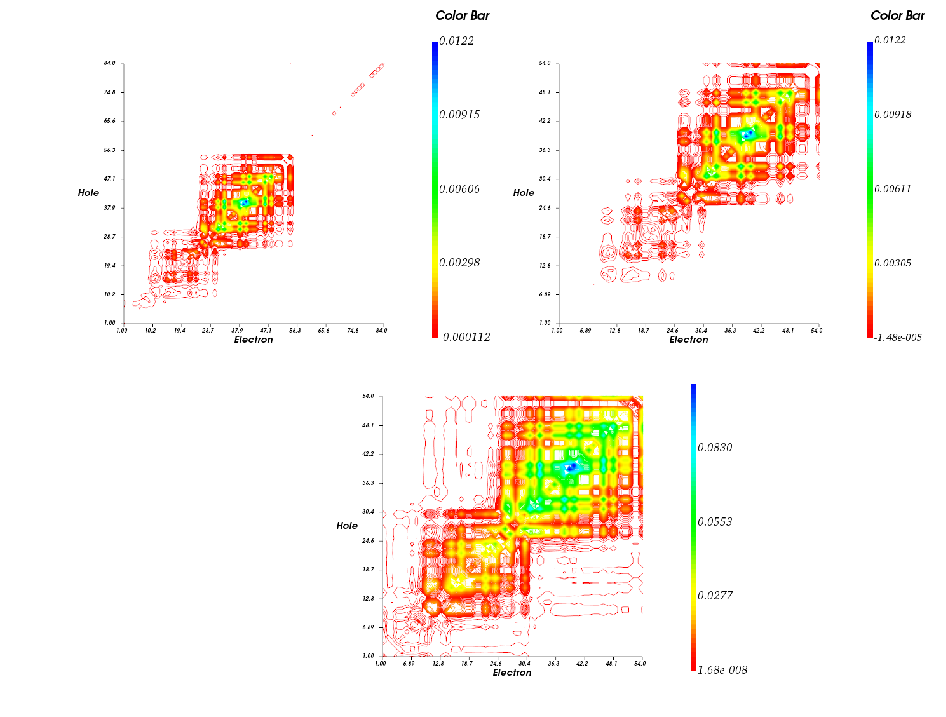}
	\end{center}
	\vspace{-0.5cm}
	\caption{CTNMs of the first excited state. Top left: CTNM. Top right: CTNM for heavy atoms only. Bottom: SCTNM for heavy atoms only.
		\label{CTNM-1}
	}
\end{figure}

\begin{figure}[H]
	\begin{center}
		\includegraphics[width=1.0\textwidth]{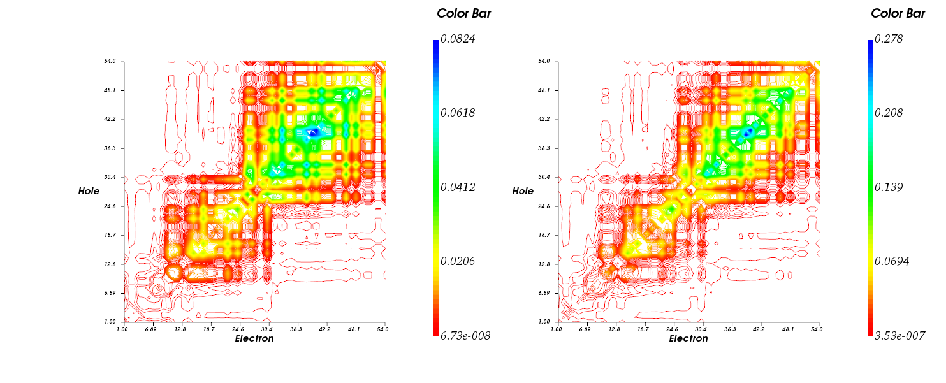}
	\end{center}
	\vspace{-0.5cm}
	\caption{Contracted Transition Density Matrices of the first excited state. Left: Method~1 (equation \ref{tdmc-1}). Right: Method 2 (equation \ref{tdmc-2}). Both of them are for heavy atoms only.
	}
	\label{CTDM-1}
\end{figure}

\begin{figure}[pth]
	\begin{center}
		\includegraphics[width=1.0\textwidth]{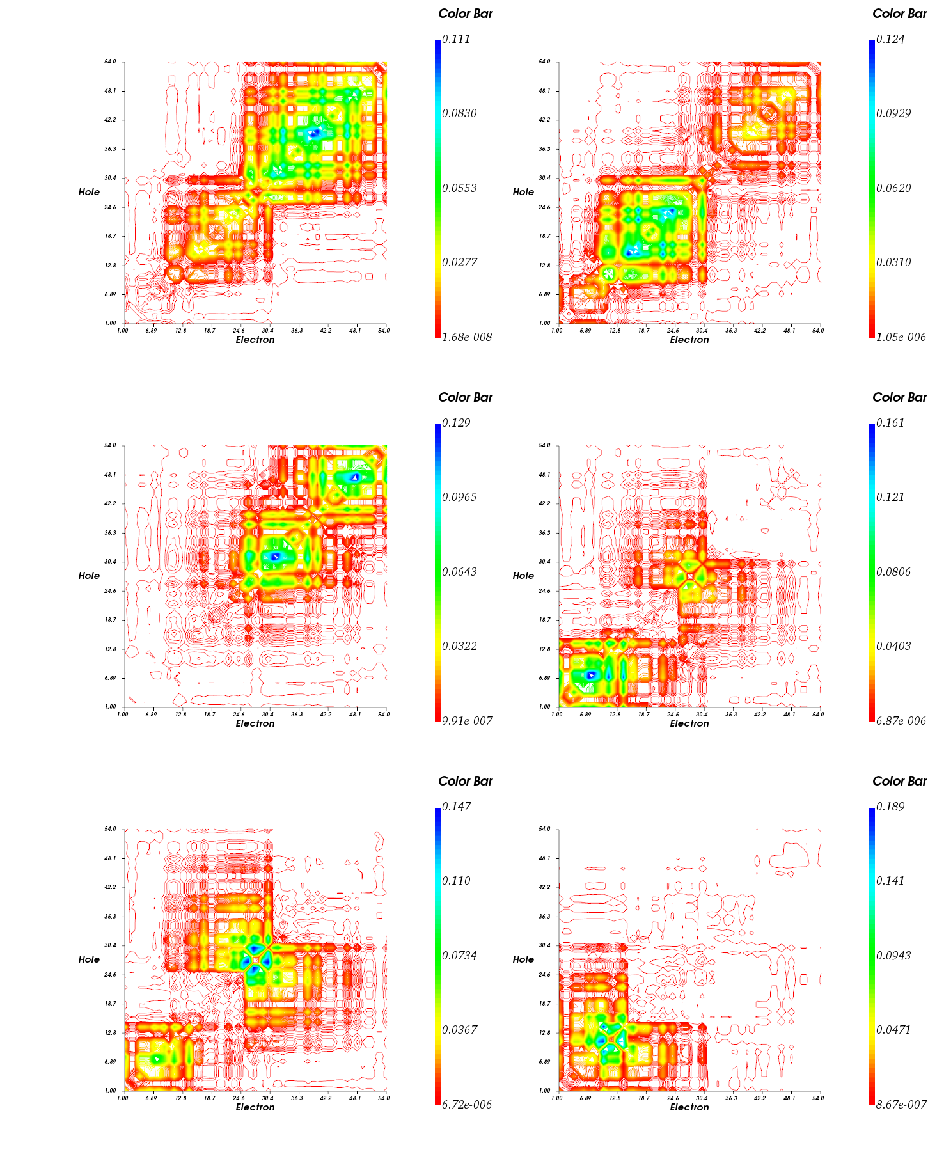}
	\end{center}
	\vspace{-0.5cm}
	\caption{SCTNMs for the first six excited states.
		\label{CTDMM1-1}
	}
\end{figure}

The overlap matrix, state Rho-CI density matrices, TDMs and CI-coefficients are extracted from rwf dumped files by GPView. With overlap matrix and TDM for each excited state, the CTDM and CTNM can be calculated. This work is much more efficient and convenient than generating 3D cubes. CTDMs and CTNMs are then displayed as contours (See Fig.~\ref{CTNM-1}, \ref{CTDM-1} and \ref{CTDMM1-1}). To get reasonable results, the nuclei have to be properly ordered. As shown in Fig.~\ref{molecule}, we sort the heavy atoms from the two-ring segment to the four-ring segment. Since hydrogen atoms play minor role in electronic excitations in such system, we leave them to the end.

%----------------------------------------------------------------------
%----------------------------------------------------------------------

%----------------------------------------------------------------------
%----------------------------------------------------------------------

The X and Y directions of contour plots represent positions of electrons and 
holes in terms of labels for nuclei.
From Fig.~\ref{CTNM-1}, we find that electronic transitions mainly occur on four-ring segment. The transitions on three-ring segment are also observed, however, they are minor. From Fig.~\ref{CTNM-1} top left, it can be seen that the electronic transitions from or to hydrogen atoms can be neglected. Therefore, in the rest of this manuscript, we will plot matrices for heavy atoms only. Contour plots of CTDMs (Fig.~\ref{CTDM-1}) and SCTNMs (Fig.~\ref{CTNM-1} bottom) give similar results. Since the sum of elements  in CTNM is close to 1 and its definition is analogous to Mayer's bond order, we suggest to use SCTNMs for displaying electronic modes. However, we can still use CTDMs if overlap matrix is not available.

%----------------------------------------------------------------------
%----------------------------------------------------------------------

The electronic modes for the first six excited states are displayed in the Fig.~\ref{CTDMM1-1}. It has been found that all of them are excitonic states, since electronic modes are symmetric and peaked along diagonal. The first and third excited states are spatially localized on four-ring segment, where zero and one node are observed. From the Exciton Scattering perspective \cite{Wu2006-NP-1}, they are two states in the first exciton band that reside on the four-ring segment. If the segment is longer, states with more than one nodes will be observed too. The second and fifth figures describe electronic transitions on the three-ring segment, while the fourth and sixth figures correspond to transitions on the two-ring segment.

%----------------------------------------------------------------------
%----------------------------------------------------------------------

The above discussed TDM based analysis has been demonstrated to be a powerful tool to investigate electronic excitations in large conjugated molecules \cite{Tretiak2002-CR-1,Wu2006-NP-1}.
Especially, in the Exciton Scattering methodology \cite{Wu2008-PRL-1,Li2013-JCP-1}, the electronic excited states belonging an exciton band are collected by viewing contour plots
of CTDMs or SCTNMs. 
However, we also need other analysis methods to get a comprehensive understanding of electronic excitations. 
In the rest of this section, we will talk about the real space wave function based analysis in GPView.

%----------------------------------------------------------------------
\begin{figure}[pth]
	\begin{center}
		\includegraphics[width=0.8\textwidth]{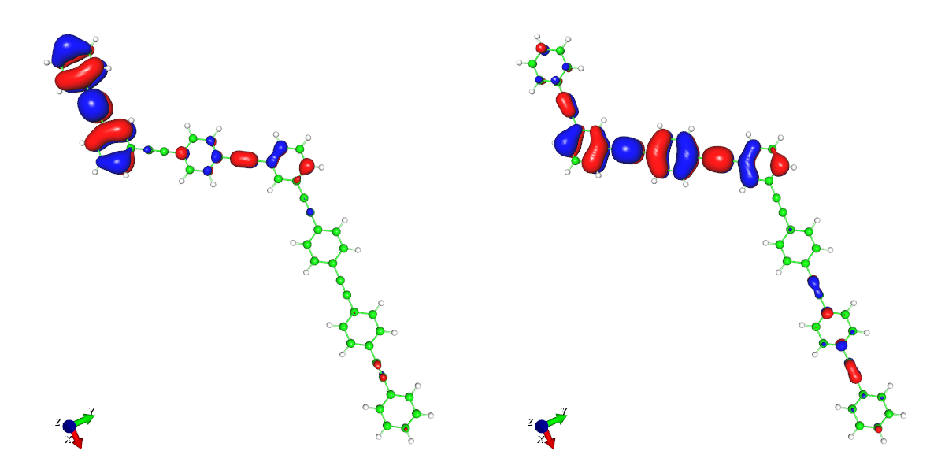}
		\includegraphics[width=0.8\textwidth]{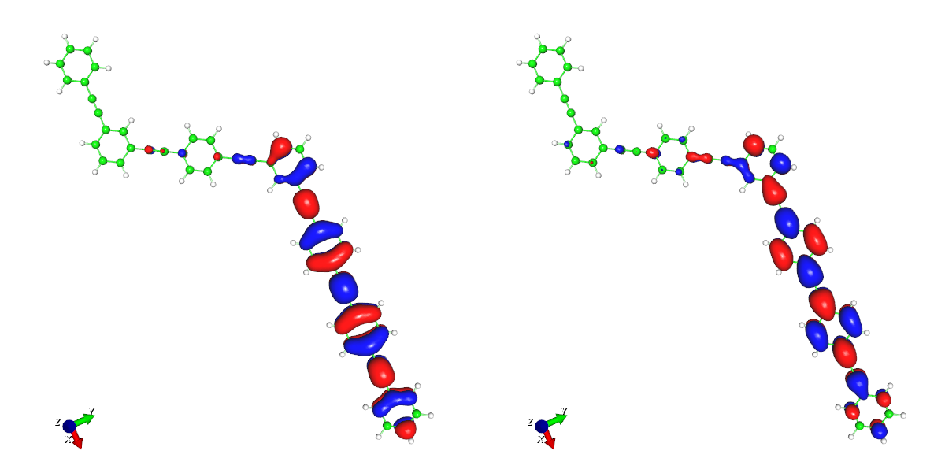}
		\includegraphics[width=0.8\textwidth]{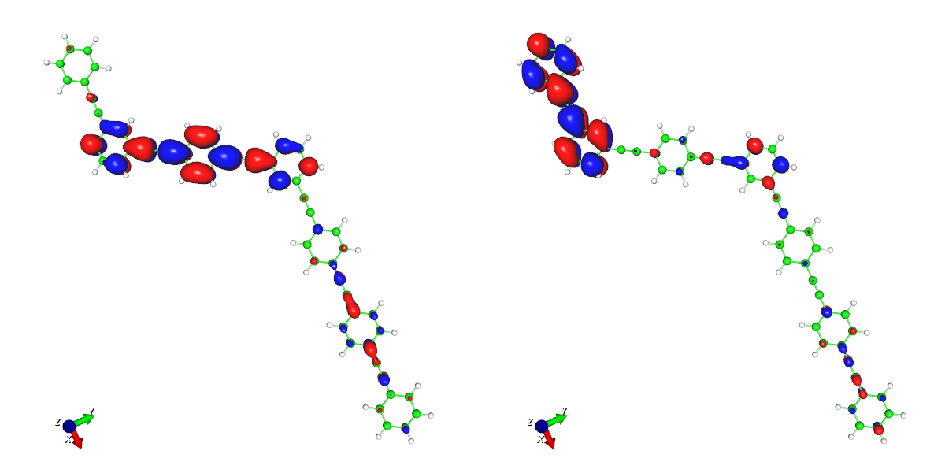}
	\end{center}
	\vspace{-0.5cm}
	\caption{Molecular Orbitals (from HOMO-2 to LUMO+2). The orbital energy are -0.2655, -0.2521, -0.2477, -0.0305, -0.0230, -0.0031 a.u, respectively.
	}
	\label{fig:c5-MO}
\end{figure}
%----------------------------------------------------------------------

Parameters of the real space wave functions are stored in Gaussian fchk files. They can be extracted and used to calculate molecular orbitals and electron density. 
We first use GPView to generate molecular orbitals range from
HOMO-2 to LUMO+2, where HOMO and LUMO represent highest occupied molecular orbital and lowest unoccupied molecular orbital, respectively. 
It can be seen from Fig.~\ref{fig:c5-MO} that HOMO and LUMO  are spatially localized on the four-ring segment, while HOMO-1 and LUMO+1 are on the three-ring segment, and HOMO-2
and LUMO+2 are on the two-ring segment.

Focusing on a single electronic excited state, we can calculate parameters for NTOs, NOs and NDOs by GPView. 
These calculations rely on Gaussian fchk files, as well as overlap matrices, state density matrices, and transition density matrices, which have been extracted from rwf files.
The parameters can be further used to generate molecular orbitals (NTOs and NDOs) and electron density (hole-particle density, transition density and detachment-attachment density).

\begin{figure}[pth]
	\begin{center}
		\includegraphics[width=0.8\textwidth]{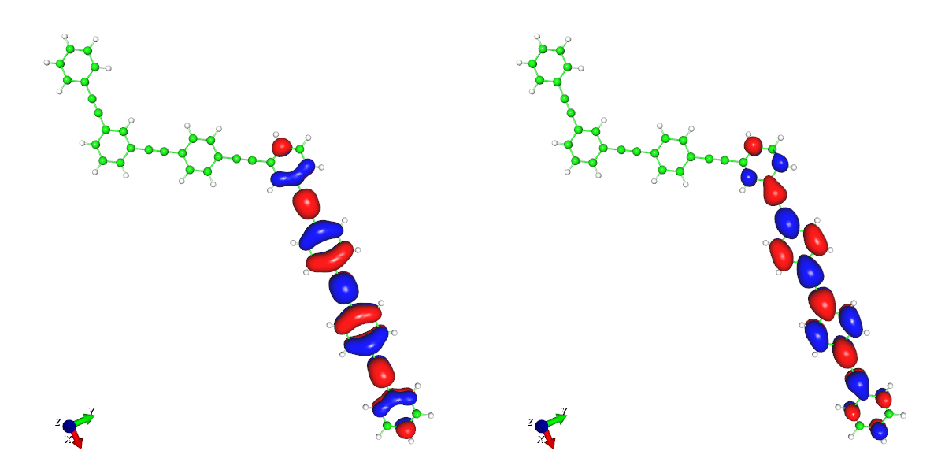}
		\includegraphics[width=0.8\textwidth]{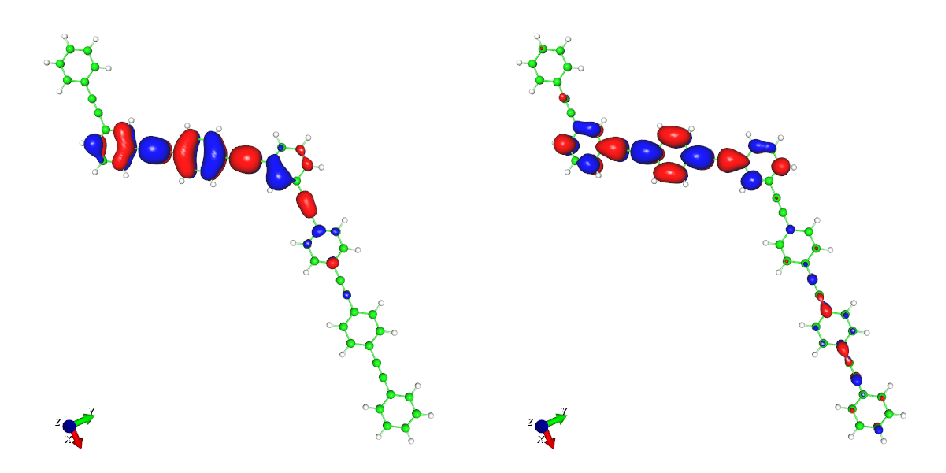}
		\includegraphics[width=0.8\textwidth]{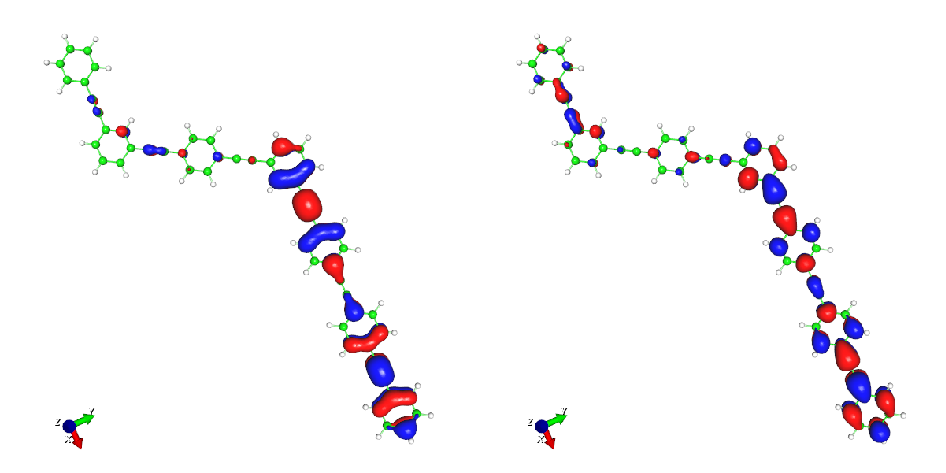}
	\end{center}
	\vspace{-0.5cm}
	\caption{Natural Transition Orbitals. Top: $\sqrt{\lambda_1}=0.8917$. Middle: $\sqrt{\lambda_2}=0.2970$. Botton: $\sqrt{\lambda_3}=0.2845$.
	}
	\label{fig:c5-nto}
\end{figure}

Natural transition orbitals of the first excited state are shown in Fig~\ref{fig:c5-nto}. 
The total weight of three pairs of NTOs in the figure exceeds 0.96, thus contributions of others can be neglected in the analysis ($\lambda_i>\lambda_{i+1}$, $i=1,2,\cdots$).
Hole and particle NTOs corresponding to $\lambda_1$ and $\lambda_3$ are spatially localized on the four-ring segment, while the pair of NTOs associated with $\lambda_2$ are on the three-ring segment. 
Since $\lambda_1/\lambda_2=9.0141$, the electronic excitations on the three-ring segment can not be ignored. 
From contour plots of SCTNMs for the first excited state (see Fig.~\ref{CTDMM1-1}), we also observe transitions on the three-ring segment.
%$\lambda_1/\lambda_3=9.8237$, 
%----------------------------------------------------------------------
%----------------------------------------------------------------------

\begin{figure}[pth]
	\begin{center}
		\includegraphics[width=0.8\textwidth]{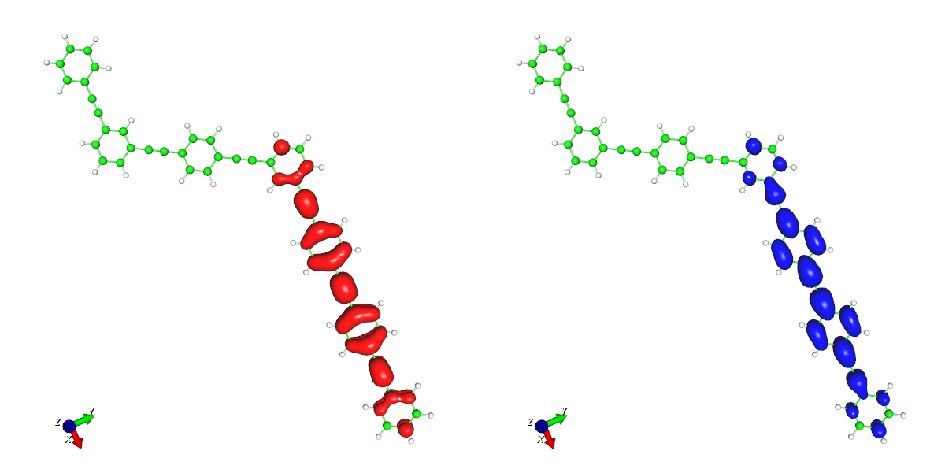}
		\includegraphics[width=0.8\textwidth]{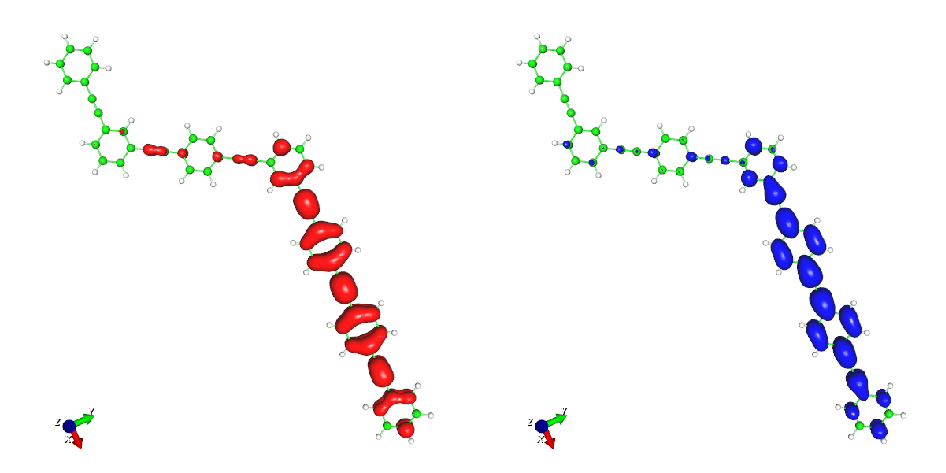}
		\includegraphics[width=0.8\textwidth]{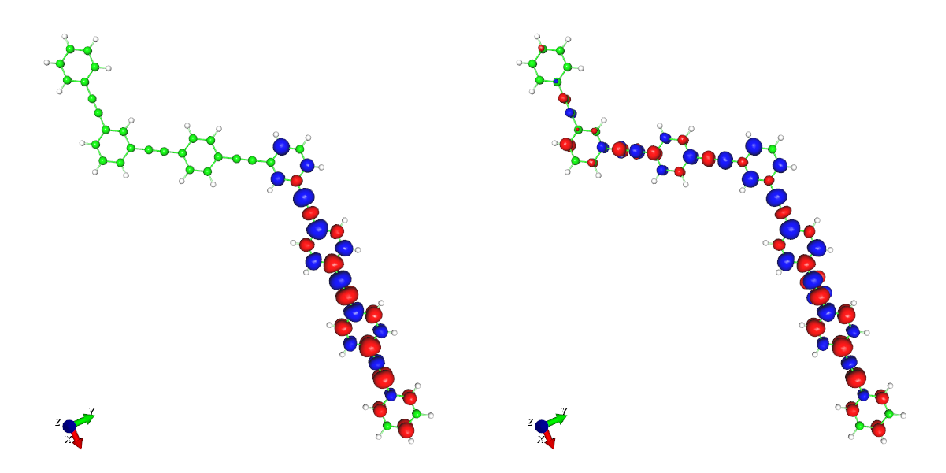}
	\end{center}
	\vspace{-0.5cm}
	\caption{Hole, particle and transition density. Top: hole-particle density calculated by NTOs with the highest weight. Middle: hole-particle density calculated by all NTOs with non-zero eigenvalues. Botton: Transition density calculated by NTOs with the highest weight and all non-zero eigenvalues, respectively.
		\label{HPTD_1}
	}
\end{figure}

The hole, particle and transition density (see Fig.\ref{HPTD_1}) are calculated as well with parameters of NTOs. 
They can be obtained in two ways. 
The first way considers contributions of all NTOs, which gives us accurate results.
The other only takes the pair of NTOs corresponding to $\lambda_1$ into consideration.
It is more efficient for large systems, however, some features may not be captured.
For example, in Fig~\ref{HPTD_1}, the hole and particle density obtained by the second way
do not show transitions on the three-ring segment.
Therefore, only if $\lambda_1\gg \lambda_2$, we can use one pair of NTOs to calculate the densities.
%----------------------------------------------------------------------
%----------------------------------------------------------------------

\begin{figure}[pth]
	\begin{center}		
		\includegraphics[width=0.8\textwidth]{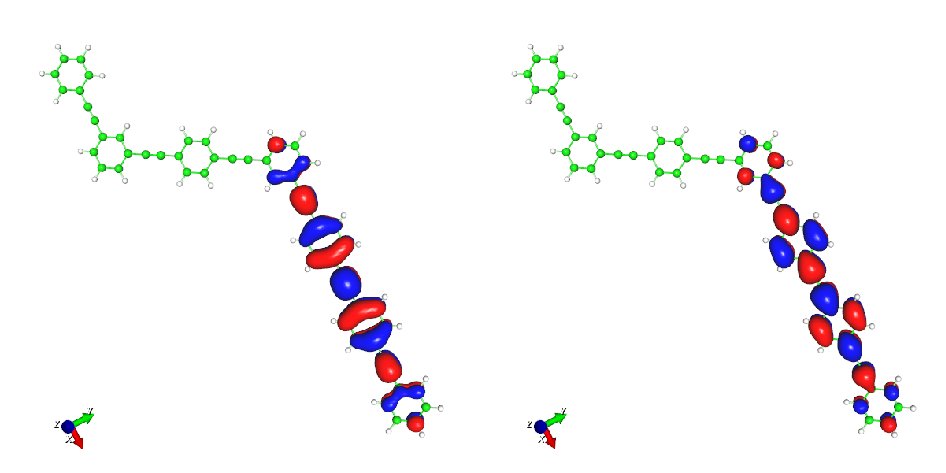}
		\includegraphics[width=0.8\textwidth]{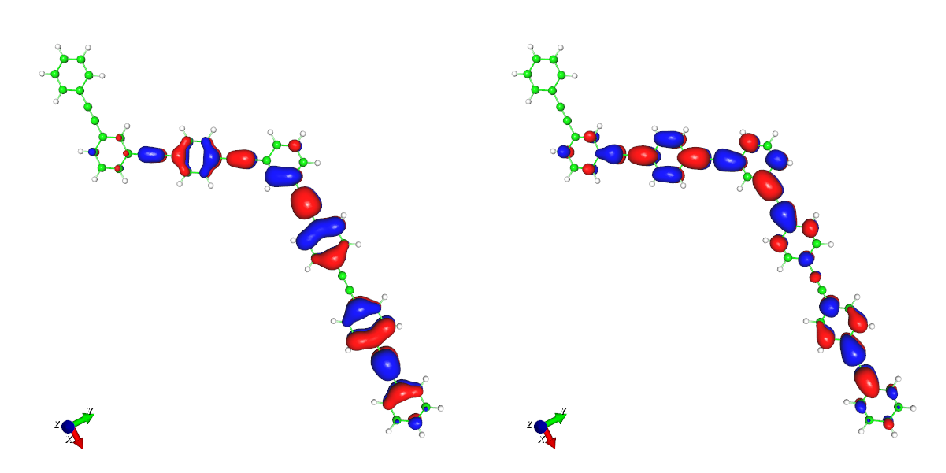}
		\includegraphics[width=0.8\textwidth]{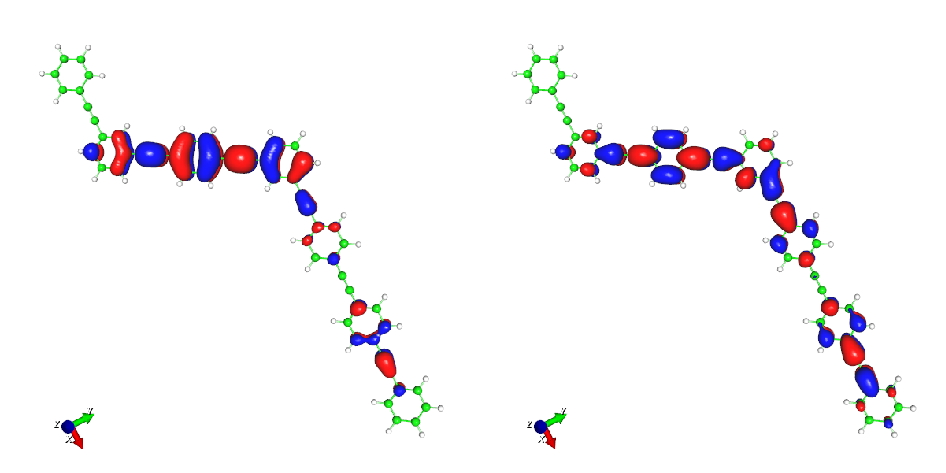}
	\end{center}
	\vspace{-0.5cm}
	\caption{Natural Difference Orbitals (CI density matrix). Top: $d_i=0.7971, a_i=0.7973$. Middle: $d_i=0.0912, a_i=0.0903$. Botton: $d_i=0.0877, a_i=0.0870$.
		\label{NDO_CI}
	}
\end{figure}

\begin{figure}[pth]
	\begin{center}		
		\includegraphics[width=0.8\textwidth]{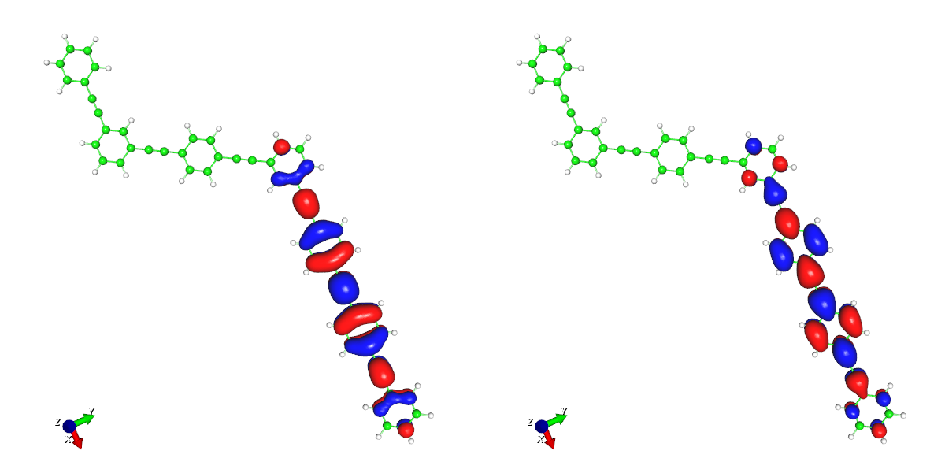}
		\includegraphics[width=0.8\textwidth]{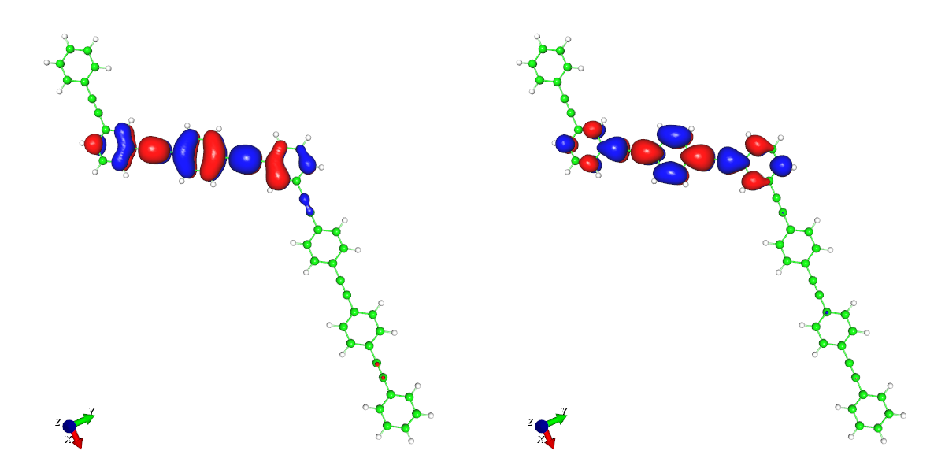}
		\includegraphics[width=0.8\textwidth]{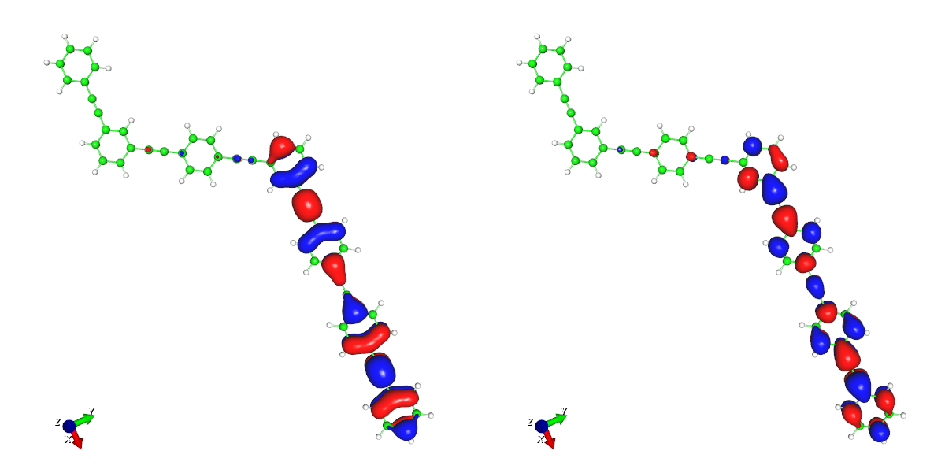}
	\end{center}
	\vspace{-0.5cm}
	\caption{Natural Difference Orbitals (Rho-CI density matrix). Top: $d_i=0.7961, a_i=0.7961$. Middle: $d_i=0.0882, a_i=0.0882$. Botton: $d_i=0.0810, a_i=0.0810$.
		\label{NDO_RhoCI}
	}
\end{figure}

Both NOs and NDOs are related with the CI density matrix of an excited state.
From quantum chemistry calculation by Gaussian 09, we obtain two different density matrices, i.e.
CI density matrix and Rho-CI density matrix. The latter is so called un-relaxed density.
Since the NDOs may contain non-intuitive contributions deriving from orbital relaxation effects \cite{plasser2014new1}, which makes them different from NTOs, we also test the NDOs calculated
by adopting Rho-CI density matrices. Three pairs of NDOs by using CI and Rho-CI density matrices are shown in Fig.\ref{NDO_CI} and \ref{NDO_RhoCI}, respectively.

\begin{figure}[pth]
	\begin{center}
		\includegraphics[width=0.8\textwidth]{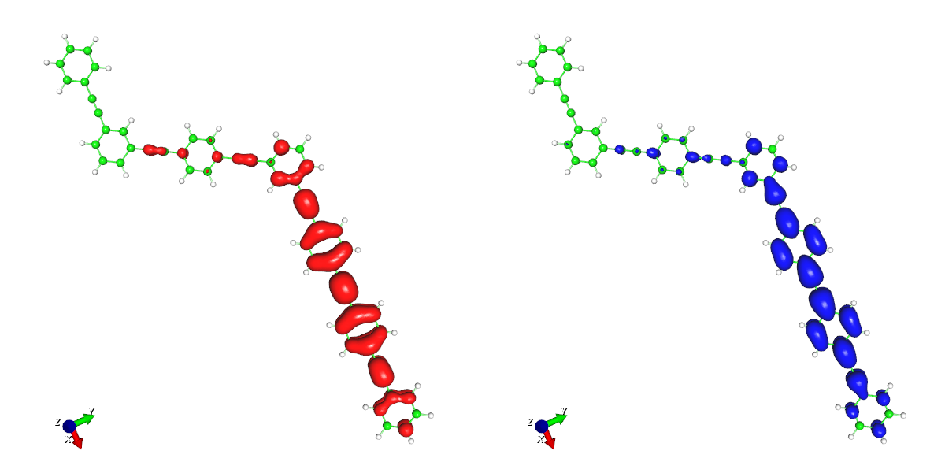}
	\end{center}
	\vspace{-0.5cm}
	\caption{Detachment-Attachment Density (Rho-CI).
		\label{DAD_1}
	}
\end{figure}

In Fig.~\ref{NDO_CI}, the primary pair of NDOs is localized on the four-ring segment, while the second and third pairs are on both four-ring and three-ring segment. However, in Fig.~\ref{NDO_RhoCI}, the primary and the third pair are on the four-ring segment, and the second pair is on the three-ring segment.
Compare Fig.~\ref{fig:c5-nto} with \ref{NDO_CI} and \ref{NDO_RhoCI}, we conclude that
NDOs from Rho-CI density matrix agree well with NTOs, regardless phases of wave functions.
This result has also been verified by a series of calculations for molecule pyridine in
user's manual of GPView.
The overall weights is close to $1.0$, which indicates promoted electrons is close to one.
The detachment and attachment density are shown in Fig.~\ref{DAD_1}.

\begin{figure}[pth]
	\begin{center}
		\includegraphics[width=0.8\textwidth]{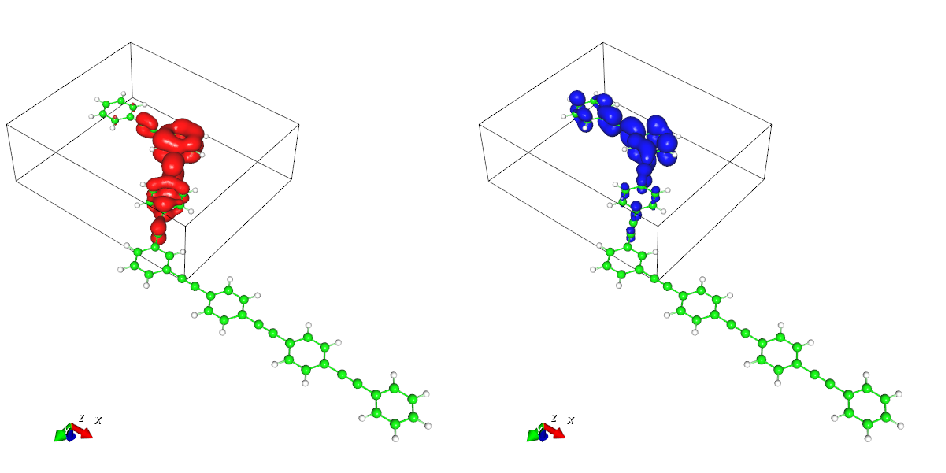}
	\end{center}
	\vspace{-0.5cm}
	\caption{Hole and particle density for the sixth excited state.
		\label{CTDM-SB-1}
	}
\end{figure}

Finally, we will introduce a convenient function that can help the module to create cube files more efficiently. It is known that calculations of grid points for orbitals and density is time consuming for large molecules. There are two ways to improve efficiency in GPView. First, we can generate low dense cubes, which means the number of points in each direction is less than before. This approach can still generate cubes with the same size, however, the quality of the figure may be coarse. Second, we can still keep the quality however generate a smaller cube (see Fig.~\ref{CTDM-SB-1}).

%----------------------------------------------------------------------
%----------------------------------------------------------------------

%----------------------------------------------------------------------
%----------------------------------------------------------------------

%----------------------------------------------------------------------
%----------------------------------------------------------------------

%----------------------------------------------------------------------
%----------------------------------------------------------------------

\section{Conclusions}
A new software is introduced in this manuscript.  
It is developed for wave function analysis and visualization purpose.
Several tools are incorporated into GPView to explore electronic structures of excited states.
TDMs are fingerprints of excited states. They can be extracted from QC calculations and then converted to CTDMs or CTNMs, that  are visualized by contour plots or color maps. 
In the ES methodology, color maps of CTDMs are widely used to classify excited states into different exciton bands, and to distinguish excitonic and charge transfer states.
For the wave function based analysis, MOs, NTOs and NDOs can be generated by GPView.
As aforementioned, electron and hole pairs are components of NTOs, while NOs
refer to individual unpaired electrons, and attachment and detachment processes are described by NDOs. 
It is also found that NDOs obtained by diagonalizing Rho-CI density matrices agree well with NTOs from plots of orbitals.
These tools provide more options to study electronic excitations for molecules.

\section{Acknowledgments}
We would like to thank Dr. Vladimir Chernyak and Dr. H. Bernhard Schlegel for
insightful discussions.

\section*{References}

\bibliography{mybibfile}

\appendix
\section{Wave function analysis}

In this section, we will explain how to use GPView to perform the wave function analysis.
In the main paper, the quantum chemistry (QC) calculations have been performed on a branched dendritic phenylacetylene (PA) based molecule, which is composed of two-, three- and four- ring linear poly(phenylene ethynylene) (PPE) units linked through meta-substitutions \cite{soler2012analysis}.
The molecule will be referred to as 1m2m3 in the rest of this paper.

\subsection{Input}
First, we create a file named 1m2m3.gjf. This is a sample Gaussian 09 input.
\begin{lstlisting}[frame=single]
%chk=1m2m3.chk
%rwf=1m2m3.rwf
%save
#p td=(nstate=10,root=1) cam-b3lyp/6-31g density=current

1m2m3

0 1
C                 17.58405600    7.72805300    0.00000000
C                 16.88519600    6.52141100    0.00000000
C                 16.88406800    8.93403900    0.00000000
C                 15.49480900    6.51708900    0.00000000
C                 15.49367400    8.93701100    0.00000000
C                 14.78135200    7.72671900    0.00000000
C                 13.35195600    7.72589100    0.00000000
C                 12.13943900    7.72454300    0.00000000
C                 10.71007800    7.72472200    0.00000000
C                 10.00216300    6.51613200    0.00000000
C                  9.99690300    8.93470800    0.00000000
C                  8.60143600    6.50481500    0.00000000
C                  8.60637100    8.92695900    0.00000000
C                  7.90665000    7.72541900    0.00000000
C                  7.88993800    5.26534800    0.00000000
C                  7.28502900    4.21417200    0.00000000
C                  6.57367900    2.97683500    0.00000000
C                  7.26430900    1.75339000    0.00000000
C                  5.16888800    2.95776400    0.00000000
C                  6.57352600    0.55172000    0.00000000
C                  4.47809000    1.75610300    0.00000000
C                  5.16876000    0.53271500    0.00000000
C                  4.45737500   -0.70465500    0.00000000
C                  3.85245700   -1.75581300    0.00000000
C                  3.14093100   -2.99517900    0.00000000
C                  3.83544100   -4.21601500    0.00000000
C                  1.74026500   -3.00661300    0.00000000
C                  3.13598500   -5.41775700    0.00000000
C                  1.03273000   -4.21549600    0.00000000
C                  1.74552800   -5.42575500    0.00000000
C                 -0.39634200   -4.21526600    0.00000000
C                 -1.60913700   -4.21644200    0.00000000
C                 -3.03634400   -4.21694500    0.00000000
C                 -3.75268700   -5.42555300    0.00000000
C                 -3.75322600   -3.00864800    0.00000000
C                 -5.13868700   -5.42592600    0.00000000
C                 -5.13922300   -3.00894100    0.00000000
C                 -5.85581400   -4.21759400    0.00000000
C                 -7.28253400   -4.21790300    0.00000000
C                 -8.49585200   -4.21809200    0.00000000
C                 -9.92263900   -4.21818800    0.00000000
C                -10.63946600   -5.42665600    0.00000000
C                -10.63958900   -3.00979200    0.00000000
C                -12.02546600   -5.42663600    0.00000000
C                -12.02558900   -3.00995300    0.00000000
C                -12.74260400   -4.21833000    0.00000000
C                -14.16960600   -4.21839500    0.00000000
C                -15.38263800   -4.21843000    0.00000000
C                -16.81155300   -4.21843400    0.00000000
C                -17.52461700   -5.42849500    0.00000000
C                -17.52461700   -3.00837300    0.00000000
C                -18.91494500   -5.42479500    0.00000000
C                -18.91494500   -3.01207100    0.00000000
C                -19.61436800   -4.21843200    0.00000000
H                 18.66813300    7.72856100    0.00000000
H                 17.42510000    5.58129800    0.00000000
H                 17.42309700    9.87466000    0.00000000
H                 14.94845800    5.58134400    0.00000000
H                 14.94650700    9.87229400    0.00000000
H                 10.54451100    5.57901900    0.00000000
H                 10.54219100    9.87066800    0.00000000
H                  8.06362000    9.86509900    0.00000000
H                  6.82345700    7.71874100    0.00000000
H                  8.34758300    1.75791300    0.00000000
H                  4.62745200    3.89604400    0.00000000
H                  7.11501100   -0.38654400    0.00000000
H                  3.39480600    1.75164700    0.00000000
H                  4.91862400   -4.20932400    0.00000000
H                  1.19786600   -2.06953500    0.00000000
H                  3.67893700   -6.35574600    0.00000000
H                  1.20029500   -6.36173500    0.00000000
H                 -3.20885600   -6.36246500    0.00000000
H                 -3.20975000   -2.07154300    0.00000000
H                 -5.68212300   -6.36305200    0.00000000
H                 -5.68306300   -2.07205400    0.00000000
H                -10.09584200   -6.36371200    0.00000000
H                -10.09606000   -2.07268200    0.00000000
H                -12.56912600   -6.36363400    0.00000000
H                -12.56933900   -2.07300800    0.00000000
H                -16.97786300   -6.36401000    0.00000000
H                -16.97786400   -2.07285800    0.00000000
H                -19.45445700   -6.36513100    0.00000000
H                -19.45445400   -2.07173300    0.00000000
H                -20.69843900   -4.21843100    0.00000000

\end{lstlisting}

\subsection{GPView analysis and visualization modules}

\begin{figure}[H]
	\begin{center}
		\includegraphics[width=0.6\textwidth]{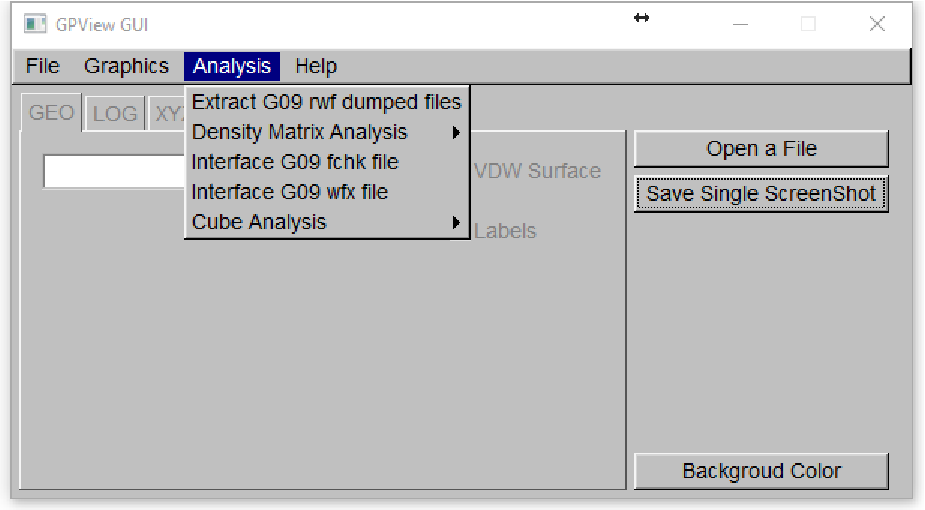}
		\includegraphics[width=0.6\textwidth]{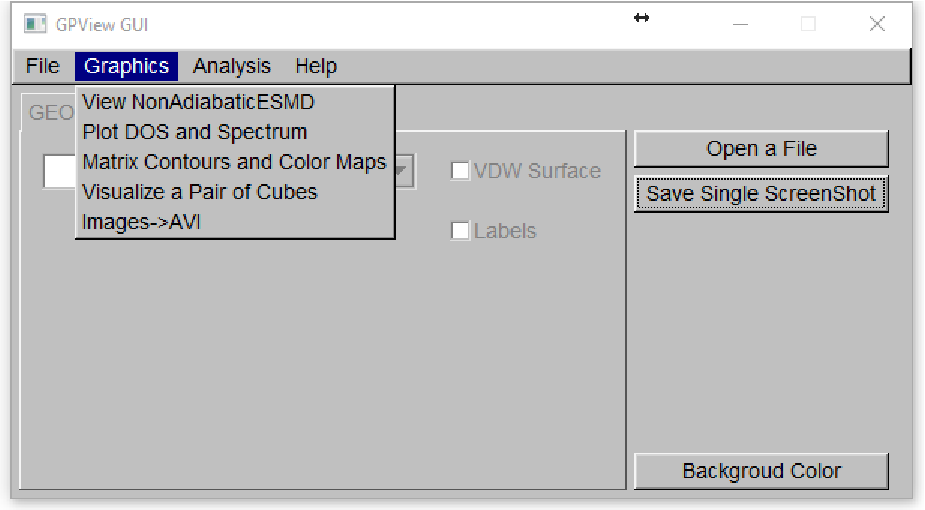}
	\end{center}
	\vspace{-0.5cm}
	\caption{GPView analysis and visualization modules.
	}
	\label{fig:modules}
\end{figure}

In Fig.~\ref{fig:modules}, we show the analysis and visualization modules of GPView.
We will introduce some functions in details in following sections.

\subsection{RWF Dump}
After the QC calculations, we have to run the following commands to dump the results in the `1m2m3.rwf' file, that is in binary format, to readable text files.
\begin{lstlisting}[frame=single]
rwfdump 1m2m3.rwf 514.txt 514R
rwfdump 1m2m3.rwf 633.txt 633R
rwfdump 1m2m3.rwf 635.txt 635R
\end{lstlisting}

The overlap matrix is stored in 514.txt file.
The state Rho-CI density matrix and one-electron transition density matrix are stored in 633.txt file.
The CI coefficients are stored in 635.txt file.
However, we can not directly use the matrices in these files, since along with them, there are meta-data and other irrelevant numbers in the files.
Therefore, we will use GPView to process these data and write them in matrix format.

\subsection{Extract RWF dumped files}

In this section, we will talk about using GPView to extract matrices from the RWF dumped files. 
First, in the analysis module (see Fig.~\ref{fig:modules}), choose
\begin{lstlisting}[frame=single]
Extract G09 rwf dumped files
\end{lstlisting}
Then, the following window will pop out.
\begin{figure}[H]
	\begin{center}
		\includegraphics[width=0.5\textwidth]{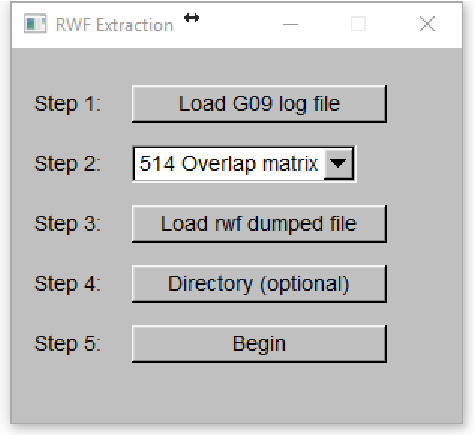}
	\end{center}
	\vspace{-0.5cm}
	\caption{Extract RWF dumped files.
	}
	\label{fig:mod-rwf-extract}
\end{figure}

In the drop-down list of step2, choose
\begin{lstlisting}[frame=single]
514 Overlap matrix
633 Excited-state CI densities
635 CIS coefficients
\end{lstlisting}
to extract 514.txt, 633.txt and 635.txt, respectively. 
The matrices are written into different files in matrix format or half-matrix format.

\subsection{Charge Transfer Number Matrix (CTNM)}
\label{sec:cal-ctnm}

\begin{figure}[H]
	\begin{center}
		\includegraphics[width=0.5\textwidth]{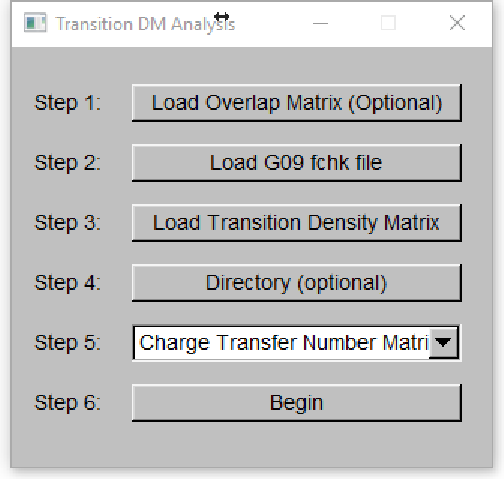}
	\end{center}
	\vspace{-0.5cm}
	\caption{One-electron Transition Density Matrix Analysis.
	}
	\label{fig:mod-1tdm}
\end{figure}

In this section, we will calculate the CTNM. 
In the analysis module, choose
\begin{lstlisting}[frame=single]
Density Matrix Analysis / 1TDM Analysis
\end{lstlisting}
Then, a new window pops out (see Fig.~\ref*{fig:mod-1tdm}).

In the drop-down list of step 5, choose
\begin{lstlisting}[frame=single]
Charge Transfer Number Matrix AO + AT + ATH + SATH
\end{lstlisting}
to calculate CTNM based on overlap matrix and transition density matrix. 
Here, AO represents Atomic Orbitals. 
Each element of CTNM-AO characterizes the transition from one atomic orbital to another.
AT represents AToms.
Each element of CTNM-AT characterizes the transition from one atom to another.
Here, CTNM-AT corresponds to CTNM in the main paper.
ATH represents Heavy AToms, i.e., atoms without hydrogen.
It is obvious that CTNM-ATH is a submatrix of CTNM-AT.
It depicts the transitions between different heavy atoms.
SATH represents Square root and Heavy AToms.
We take the square root element-wise for CTNM-ATH, then we can obtain CTNM-SATH, which corresponds to SCTNM in the main paper.

\subsection{Visualize CTNM-SATH}

In this section, we will talk about how to visualize the CTNM-SATH obtained in section \ref{sec:cal-ctnm}.
In the visualization module, choose the menu
\begin{lstlisting}[frame=single]
Matrix Contours and Color Maps
\end{lstlisting}
Then, you will see the following windows.

\begin{figure}[H]
	\begin{center}
		\includegraphics[width=0.8\textwidth]{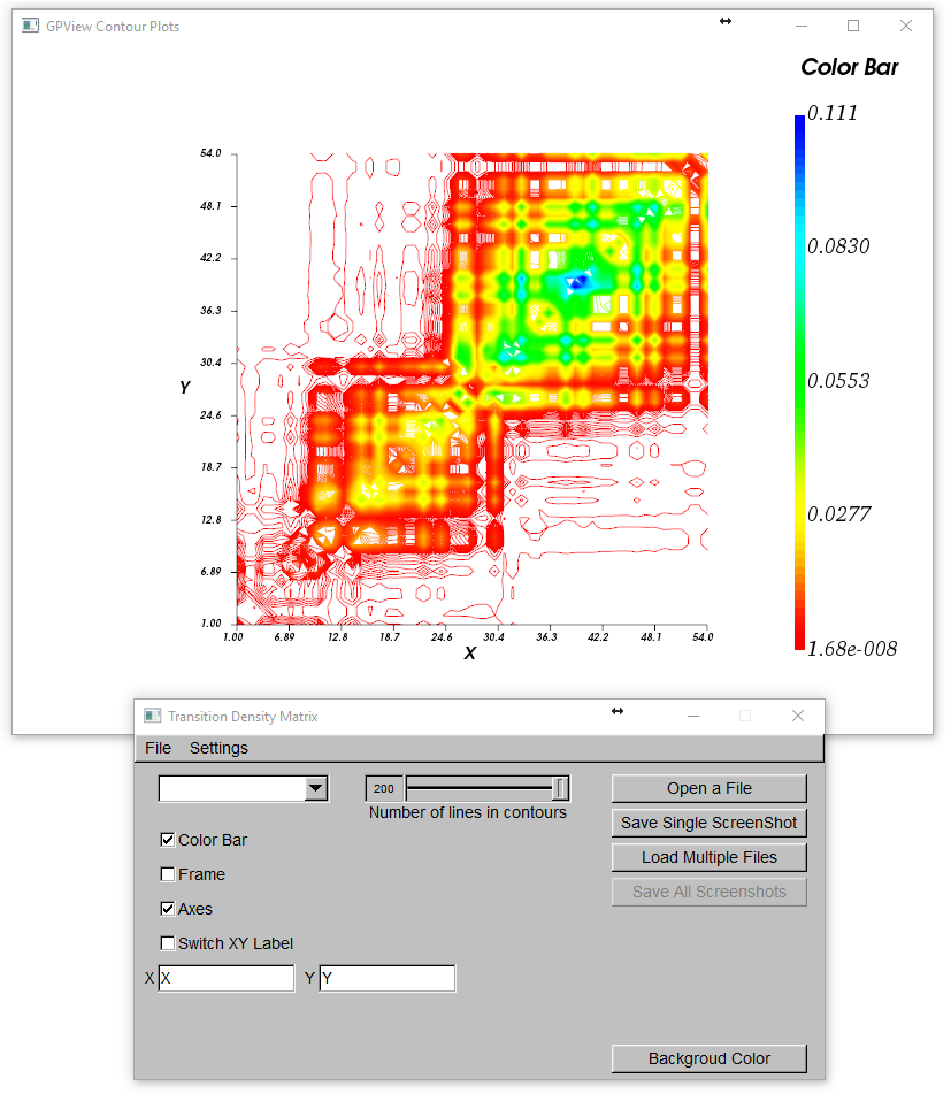}
	\end{center}
	\vspace{-0.5cm}
	\caption{Visualization of Transition Density Matrix.
	}
	\label{fig:mod-view-ctnm-sath}
\end{figure}
The CTNM-SATH can be displayed as contours or color maps.

\subsection{Molecular Orbitals}

In this section, we will show how to calculate and generate molecular orbitals and electron densities.
Suppose you need to visualize HOMO-LUMO. 
Then, by following these steps, you will get the cube files for HOMO and LUMO.

First, in the analysis module, choose
\begin{lstlisting}[frame=single]
Interface G09 fchk file
\end{lstlisting}
Then, the following window pops out.

\begin{figure}[H]
	\begin{center}
		\includegraphics[width=0.8\textwidth]{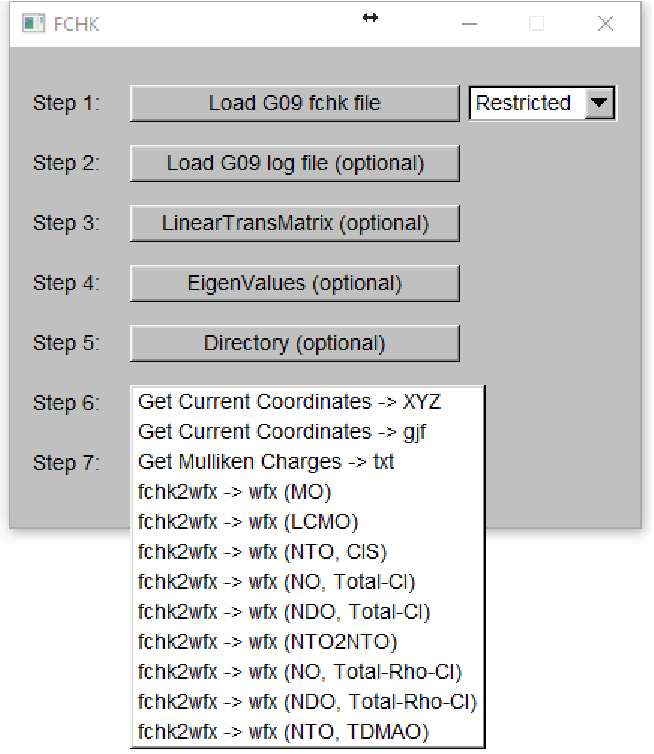}
	\end{center}
	\vspace{-0.5cm}
	\caption{Interface with fchk files.
	}
	\label{fig:mod-ana-fchk}
\end{figure}

To calculate the molecular orbitals, in Fig.~\ref{fig:mod-ana-fchk}, you can choose
\begin{lstlisting}[frame=single]
fchk2wfx -> wfx (MO)
\end{lstlisting}
Then, you will generate a wfx file, which contains the parameters of all molecular orbitals.

Now, in the analysis module, choose
\begin{lstlisting}[frame=single]
Interface G09 wfx file
\end{lstlisting}
Then, you can see the following window.

\begin{figure}[H]
	\begin{center}
		\includegraphics[width=0.8\textwidth]{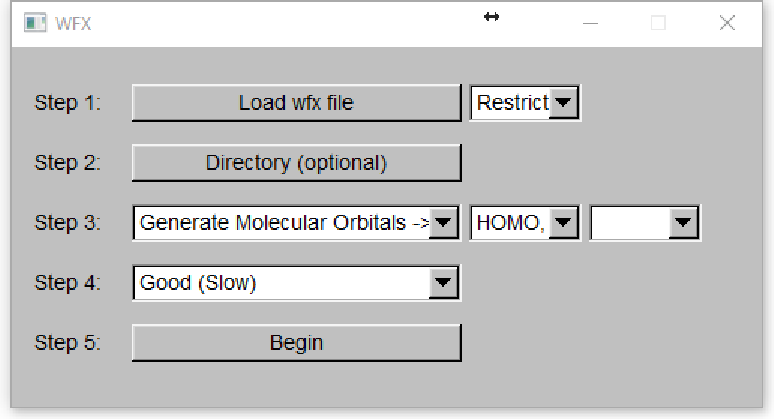}
	\end{center}
	\vspace{-0.5cm}
	\caption{Interface with wfx files.
	}
	\label{fig:mod-ana-wfx}
\end{figure}

In the drop-down list of step 3, choose
\begin{lstlisting}[frame=single]
Generate Molecular Orbitals -> cub
HOMO, LUMO
\end{lstlisting}
to generate HOMO and LUMO. Two *.cub files will be created.

In the list of step 3, you can also choose
\begin{lstlisting}[frame=single]
Generate SCF Density -> cub
Total Density
\end{lstlisting}
to generate the total SCF density.

\subsection{Natural Transition Orbitals (NTOs)}

In this section, we will talk about how to generate NTOs. 
In the Fig.~\ref{fig:mod-ana-fchk}, you have three options to calculate the parameters for NTOs.\\
Choose
\begin{lstlisting}[frame=single]
fchk2wfx -> wfx (NTO, CIS)
\end{lstlisting}
if you have already extracted 635.txt.\\
Choose
\begin{lstlisting}[frame=single]
fchk2wfx -> wfx (NTO2NTO)
\end{lstlisting}
if you have performed the NTO calculations using Gaussian 09.\\
Choose
\begin{lstlisting}[frame=single]
fchk2wfx -> wfx (NTO, TDMAO)
\end{lstlisting}
if you have extracted 514.txt and 633.txt files.

After getting the *.wfx files, you can go back to Fig.~\ref{fig:mod-ana-wfx} and choose
\begin{lstlisting}[frame=single]
Generate Molecular Orbitals -> cub
HOMO, LUMO
\end{lstlisting}
to generate the first pair of NTOs. Alternately, if you choose
\begin{lstlisting}[frame=single]
Generate Molecular Orbitals -> cub
HOMO-2, LUMO+2
\end{lstlisting}
you will get the first three pairs of NTOs.

You can also generate the hole, particle and transition density by choosing
\begin{lstlisting}[frame=single]
Generate Hole-Particle and Transition Densities (1NTO) 
-> cub
Total Density
\end{lstlisting}
if you want to use only one pair of NTOs, or
\begin{lstlisting}[frame=single]
Generate Hole-Particle and Transition Densities (AllNTO) 
-> cub
Total Density
\end{lstlisting}
if you want to use all NTOs.

\subsection{Natural Difference Orbitals (NDOs)}

The procedure of generating NDOS is very similar to that for NTOs.
Here, GPView provides two options based on total CI and total Rho-CI density matrices.
The NDOs generated by them are usually different from each other.
The ones that based on total Rho-CI density matrices are more consistent with NTOs.
In this section, we will show the procedure to generate the NDOs based on total Rho-CI density matrices, obtained from 633.txt files.
In the Fig.~\ref{fig:mod-ana-fchk}, you can choose
\begin{lstlisting}[frame=single]
fchk2wfx -> wfx (NDO, Total-Rho-CI)
\end{lstlisting}
to calculate NDOs based on total Rho-CI density matrices.
After getting the *.wfx file for each excited state, you can go to wfx interface, i.e., Fig.~\ref{fig:mod-ana-wfx}
and choose
\begin{lstlisting}[frame=single]
Generate Molecular Orbitals -> cub
1, 2, 3, 544, 545, 546
\end{lstlisting}
to generate six NDOs. 
Orbitals 1 and 546, 2 and 545 and 3 and 544 form three pairs of NDOs that have the most contributions.

If you want to get attachment and detachment density, choose
\begin{lstlisting}[frame=single]
Generate Attachment and Detachment Densities (NDOs) 
-> cub
\end{lstlisting}

\subsection{Visualize a pair of molecular orbitals}

With GPView, we can visualize the cube files obtained from previous sections.
In the visualization module, choose
\begin{lstlisting}[frame=single]
Visualize a Pair of Cubes
\end{lstlisting}
Then you will see the following windows.
You can load cube files and visualize molecular orbitals and electron densities.

\begin{figure}[H]
	\begin{center}
		\includegraphics[width=\textwidth]{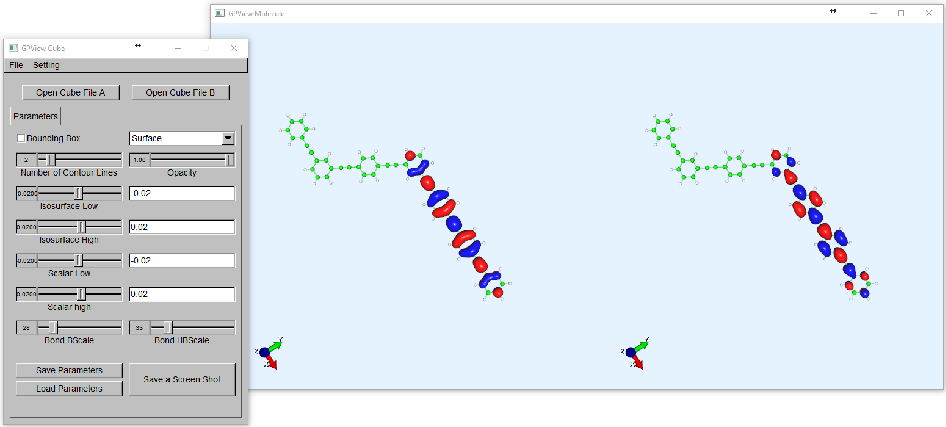}
	\end{center}
	\vspace{-0.5cm}
	\caption{Visualize a pair of molecular orbitals.
	}
	\label{fig:mod-ana-wfx}
\end{figure}

\end{document}